\documentclass[submission, Proceedings]{SciPost}
\usepackage{amsmath}
\usepackage{gensymb}
\usepackage{caption}
\usepackage[textfont=small]{caption}

\begin{document}

\begin{center}{\Large \textbf{ An improved interstitial-ice model for pure liquid water}}\end{center}

\begin{center}
J. De Poorter\textsuperscript{1}
\end{center}

\begin{center}
{\bf 1} Horatio vzw - Koningin Maria Hendrikaplein 64d, Ghent, 9000, Belgium
\\
* john.zarat@gmail.com
\end{center}

\begin{center}
\today
\end{center}


\section*{Abstract}
{\bf
The main idea of the interstitial-ice model is that liquid water consists of an intact hexagonal lattice with both vacant lattice positions and interstitial water molecules. Narten, Danford and Levy derived the model parameters from the X-ray diffraction patterns of liquid water. However, their results were counter-intuitive: their model had almost no vacancies, the interstitial concentration was temperature independent and the ice lattice was deformed resulting in extra anisotropy. To overcome these problems, we refined the model and derived the model parameters from the proton nuclear magnetic resonance frequency of pure water. Without any extra anisotropy, a significant concentration of vacancies was found (3.7\% of the lattice positions at 0$\degree$C) and the concentration of both the vacancies and interstitials increased quasi linear with temperature (0.05 M/$\degree$C). This improved model successfully explains the thermodynamic data of water and is therefore a promising candidate for a coherent and analytical water model.  }

\vspace{10pt}
\noindent\rule{\textwidth}{1pt}
\tableofcontents\thispagestyle{fancy}
\noindent\rule{\textwidth}{1pt}
\vspace{10pt}

\section{Introduction}
\label{sec:intro}
Liquids have a dual nature\cite{RN11663}. They flow, a property they share with gases but at the same time the interactions between the molecules are as strong as in solids. If one compares the latent heat of fusion of ice (6.0~kJ/mol) with the heat of sublimation of ice (51~kJ/mol at 0$\degree$C)~\cite{RN5678}, it is clear that only a small fraction of the bondings (around 10\%) is broken during the melting process. This fact is often neglected in mainstream textbooks where the motion of liquid molecules is described as free and random, "like a bunch of marbles in a bag"~\cite{RN1327}. In reality this motion is strongly limited by the bonds between the liquid molecules. 

The properties of liquid water are primarily determined by hydrogen bonds\cite{RN10889,RN7746}, also responsible for the rigid ice lattice. Several structural models for liquid water were proposed, from mixture models with different types of clustered molecules to uniformist models in which all molecules must be regarded as equivalent~\cite{RN10996,RN10889, RN9441,RN11678, RN1249}. Although, the original models were relatively simple, recent models are complex and only accessible using computational methods~\cite{RN9441,RN11678, RN1249}. Until now there is no consensus about a standard model for water. It is our opinion that a structural water model can only become relevant if it is easily extrapolated to the other fields of water physics and chemistry. We therefore reevaluated the older water models and tried to improve one of the most promising of them, the interstitial-ice model.  

Narten, Danford and Levy proposed the idea that water is ice containing a large amount of water molecules at interstitial positions~\cite{RN10885,RN11003}. They assumed that during the phase transition the ice framework undergoes minor alteration and several water molecules enter the void places enclosed by the ice lattice. These interstitial molecules are related to the 10\% larger density of water in comparison to the one of ice. This interstitial-ice model explains the X-ray diffraction patterns of liquid water~\cite{RN11667, RN11679} and quantifies the structural relaxation time and the sound absorption in both normal water and heavy water~\cite{RN9223,RN10857,RN10855}. However, the idea wasn't picked up by other groups and it became one of the many possible approaches for modelling liquid water~\cite{RN10889}. 

Reexamining the Narten model made clear that it had some hidden problems. In order to explain the X-ray diffraction data of liquid water, the ice lattice needed to be deformed, making it more anisotropic than the lattice of solid ice. Also, their ice-like lattice contained almost no vacant positions, leaving no clue for a coherent explanation for liquid water's viscosity. To overcome these problems we redeveloped the model from scratch.  Instead of using the complex X-ray diffraction patterns, we used the proton nuclear magnetic resonance frequency of pure water to find the concentration of vacancies and interstitials, an approach also used by J.C. Hindman~\cite{RN7746}. He used the same dataset to find the size of the ice clusters of the flickering-ice model for water, a two-phase model for water. The interstitial-ice model is fundamentally different from two-phase models because the interstitials are part of the lattice structure and not a separate phase. The absence of a localised second phase makes the model simple and analytical because it is independent from parameters like the distribution of the second phase, the size of the second-phase clusters, the dynamics near the phase surface, etc.  

The results of our analysis are presented in this paper. Not only will the new model solve the hidden problems, it has also the potential to become a simple and didactical way to introduce the water structure to a wider public. 

\section{The interstitial-ice model }

\subsection{The main idea}

\begin{figure*}
\centering{\includegraphics[width=130mm]{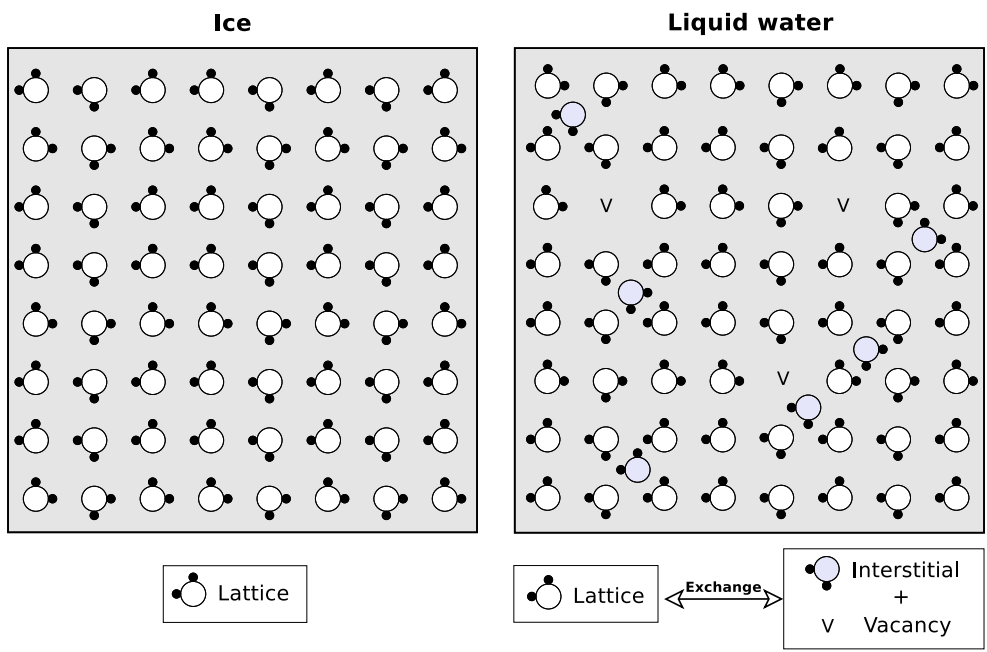}}
\caption{A 2D representation of fixed volume of ice and water. The liquid water box contains more interstitial water molecules than vacant positions explaining the larger density of liquid water. The lattice water molecules may jump to an interstitial position leaving a vacancy behind and vice versa an interstitial and vacancy may combine to a lattice water molecule. }
\label{fig:Model} 
\end{figure*}

The most common crystalline phase of ice, ice I$_h$, is hexagonal. Each water molecule is fixed in the ice crystal structure and surrounded by 4 nearest neighbours located at the corners of a tetrahedron~\cite{RN10913}.  A 2D representation of the ice lattice can be be found in left part of Fig.~\ref{fig:Model}. The ice lattice is an open structure. At 0$^{\circ}$C, only 50.9~moles of water molecules are found in one liter of ice~\cite{RN5678}. In between the lattice positions there is enough space for an extra water molecule (the so-called interstitial water molecules). In the hexagonal ice lattice there is 1~interstitial position per 2~lattice positions~\cite{RN9223}, so at 0$^{\circ}$C there are 25.5 M of potential positions for interstitials. However, the amount of interstitials in ice is negligible (only 2.8 10$^{-6}$ of the crystal sites at 0$\degree$C~\cite{RN10943}). 

The main ideas of the interstitial-ice model for liquid water are visualised in the right part of Fig.~\ref{fig:Model}. Liquid water has an intact ice-like lattice but there are two major differences with the normal ice structure. First, there is a significant concentration of empty lattice positions, the so-called vacancies. Second, an even larger concentration of interstitial positions is filled with water molecules. One liter of liquid water at 0$\degree$C contains 55.5~moles of water molecules~\cite{RN5678}. This larger density corresponds to more interstitials than vacancies present in the water lattice (see Fig.~\ref{fig:Model}).  

The quantity $n_I$ is defined as the particle density of interstitials, $n_V$ as the particle density of the vacancies, $n_{o}$ as the density of the possible positions in the ice-like lattice and $n_s = n_o-n_V$ as the density of lattice positions filled with water molecules. Mass conservation relates the density of water molecules in liquid water $n_w$ to the other densities 
\begin{equation}
n_w = n_s + n_I  = n_{o} - n_V + n_I. \label{eq:conc}
\end{equation} 

\subsection{The implementation of Narten, Danford and Levy}

The X-ray diffraction patterns of pure water depend on the spatial arrangement of the water molecules. If a water model contains the average position of the water molecules, the expected diffraction spectrum could be calculated and compared to the data. Narten, Danford and Levy tested several, at that time, common water models and found that only the interstitial-ice model gave a good fit for both large-angle and small-angle X-ray scattering~\cite{RN11003}. 

To understand how the fitting was performed, it is important to know that there are two nearest-neighbour distances in an hexagonal ice lattice~\cite{RN10943, RN10877}.  There is $r_{oo}^{ab}$ which is the nearest-neighbour distance between the water molecules in the plane of the hexagon, also called the ab plane, and there is $r_{oo}^{c}$ which is the nearest-neighbour distance in the c direction perpendicular to the hexagons. Differences between these distances makes the ice lattice anisotropic. In solid ice, there is only a small anisotropy of 1.2\% which is temperature independent~\cite{RN10943}.

Narten, Danford and Levy had to introduce some constraints to the interstitial-ice model to allow them to fit the model to the experimental data~\cite{RN10885}. One of these constraints was that the interstitial molecules are located at the triad axis of the hexagonal structure. This way, the position of these interstitials could be described with only one parameter reducing the complexity of the fit. They also did allow the lattice positions to be vacant. 

The interstitial-ice model parameters were fitted to the X ray diffraction patterns for 5 temperatures between 0 and 100$\degree$C. There was only a good fit if the anisotropy of the lattice was not fixed. So, $r_{oo}^{ab}$ and  $r_{oo}^{c}$ were optimised independently. The results are found in Table~\ref{Table5}. 

\begin{table}
\caption{The parameters derived by Narten, Danford and Levy from X-ray patterns~\cite{RN10885}. Notice that they found a significant anisotropy between the intermolecular distances in the hexagonal ab plane ($r_{oo}^{ab}$) and the same distances in the c direction ($r_{oo}^{c}$).   } 
\label{Table5}
\begin{tabular}{| c ||  c |  c |  c | c | c |c |}
\hline
T ($^{\circ}$C)& $r_{oo}^{ab}$ (\AA)& $r_{oo}^{c}$ (\AA)&$ n_o$ (M)& $ n_V$ (M)&$n_I$ (M)&anisotropy (\%)\\
\hline 
\hline 
4 & 2.91 & 2.79 & 45.2 & 0.0 & 10.3 & 4.1 \\
25 & 2.94 & 2.78 & 44.2 & 0.0 & 10.9 & 5.4 \\
50 & 2.96 & 2.74 & 44.0 & 0.0 & 10.5 & 7.4 \\
75 & 2.99 & 2.74 & 43.0 & 0.0 & 10.7 & 8.4 \\
100 & 3.02 & 2.76 & 41.9 & 0.5 & 11.3 & 8.6 \\
\hline 
\end{tabular}
\par
\end{table}

The fitted parameters reflect three clear trends. First, there are no vacant positions. Only at high temperatures (above 75 $\degree$C), vacancies are detectable. Second, the concentration of the interstitials remains more or less constant over the whole temperature range. Third, there is a significant anisotropy between the intermolecular distances in the hexagonal ab plane ($r_{oo}^{ab}$) and the c direction ($r_{oo}^{c}$), varying from 4.1\% at 4$\degree$C increasing to 8.6\% at 100$\degree$C (see Table~\ref{Table5}). 

All of these trends are contra-intuitive. The absence of significant concentrations of vacancies makes the lattice rigid, leaving no clue for a coherent explanation for liquid water's viscosity. It is also unclear why the number of interstitials is not increasing with temperature and what is the reason for the temperature-dependent anisotropy of the ice structure? It is clear that the X-ray diffraction data is sensitive to the exact location of the interstitials. In order to avoid this problem, we first refined the interstitial-ice model and used alternative data sources for the fitting of the model parameters. 

\subsection{Refinement of the model}

\begin{figure*}
\centering{\includegraphics[width=90mm]{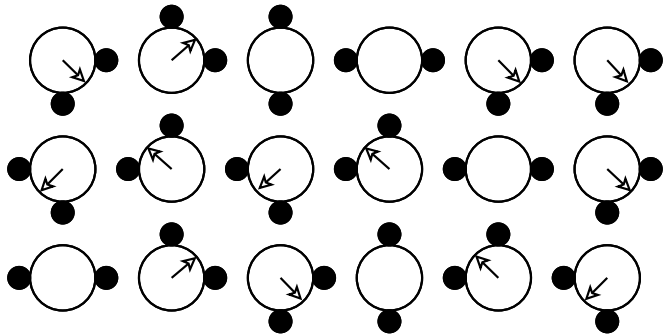}}
\caption{Schematic 2D representation of the water lattice. The dipole moment of the individual water molecules $\mathbf{p_o}$ is indicated with small arrows. Some water molecules do not have a dipole moment in the 2D plane, reflecting the 6 possible orientations the water molecules have in a hexagonal lattice~\cite{JDP1}. The local electric field around these dipoles will vary from cavity to cavity and are not symmetric around the center of the cavities.}
\label{fig:2Dmodel6}
\end{figure*}

The interstitials play a central role in interstitial-ice model. While each lattice water molecule is connected to the lattice with 4 hydrogen bonds, the interstitial water molecules are assumed to be not connected to the lattice by hydrogen bonds (\em assumption~1\em). However, this doesn't mean that they are completely free and also not that they are located at the triad axis of the hexagonal structure. Water molecules have a permanent dipole moment inducing strong local fields inside the cavities of the hexagonal lattice. The interstitials will certainly interact with these fields. Also, these fields vary from cavity to cavity due to the random orientations of the water molecules within the Bernal-Fowler ice rules~\cite{JDP1}. Even in simplified 2D representation of an ice lattice (Fig.~\ref{fig:2Dmodel6}) the different orientations of the dipole moments break the central symmetry (or the axial symmetry in the case of 3D) of the electric field inside the cavities. It is much more likely that the interstitials are not in the central triad axis and that they are bonded to the lattice structure by Van de Waals interactions. 

It is also worthwhile to have a closer look to the phase transition from a solid lattice to the liquid lattice. During the heating of a solid mono-crystalline ice crystal, parts of the ice crystal will heat up until 0$^{\circ}$C and start to transform. The  lattice structure is now sufficiently open to allow a critical amount of interstitial molecules and vacancies to be formed, significantly more than at lower temperatures. The vacancies weaken the lattice structure, allowing to expand even more. This positive feedback results in a high density of interstitials and vacancies in the transforming ice parts (around 8 M, as will be shown later). So, the lattice of the liquid water will have a lower density than the solid ice lattice (\em assumption 2\em). Because the expansion of the lattice is small compared to the mean distance between the water molecules, it is not visualised in Fig.~\ref{fig:Model}. 

Due to the uneven expansion between the transforming ice parts and the rest of the ice crystal, pressure gradients will build up. These gradients drive vacancies to the borders of the liquid ice parts disconnecting them from the solid ice crystal. The liquid grains will merge together with the existing liquid fraction. 

During the phase transition, the lattice expansion decreases the water density in contrast to the larger water density that is found experimentally. It is important to realise that the forming of more interstitials does not solve this problem. If a water molecule jumps from a lattice site to an interstitial site, it leaves a vacancy in the lattice. So the water density is not changed by this. Only if the vacancies leave the lattice structure at the surface, the density may increase. It is assumed that this volume shrinkage only happens during the phase transition when the ice lattice is broken into smaller liquid grains. A significant part of the vacancies are leaving the water structure during the phase transition at 0$\degree$C, so their density in liquid water at 0$^{\circ}$C, noted as $n_{Vo}$, is smaller than the density of the interstitial water molecules $n_{Io}$ at the same temperature (\em assumption~3\em). 

We will show further that the value of $n_{Vo}$ is not negligible as in the Danford model. For the improved model a value of 1.8 M is found, many orders of magnitude larger than in solid ice~\cite{RN10913}. Moreover, $n_V$ will increase with temperature because for every interstitial formed at higher temperatures, a vacancy remains in the lattice. This correspondence can be quantified as 
\begin{equation}
n_{V} - n_{Vo}=  n_{I} - n_{Io} .  \label{eq:nVnI}
\end{equation} 
As a consequence, Eq.~\ref{eq:conc} can be rewritten as
\begin{equation}
n_w = n_{o} + n_{Io} - n_{Vo}, \label{eq:conc2}
\end{equation} 
showing that within the interstitial-ice model changes of the water density are completely induced by changes in the lattice density $n_o$. 

A fundamental quantity of the lattice is $r_{oo}$, the distance between two oxygen atoms of neighbouring hydrogen bonded water molecules. In a tetrahedral lattice (or a hexagonal lattice with a negligible anisotropy), the density $n_o$ relates to $r_{oo}$ as~\cite{RN10878}
\begin{equation}
n_o = \frac{ 3 \sqrt{3} }{8r_{oo}^3}. \label{eq:no}
\end{equation}
At a temperature of 25$^{\circ}$C, the distance $r_{oo}$ can be extracted out of the X-ray diffraction spectra with a 1\% accuracy, 2.80 \AA\cite{RN1132}. Using Eq.~\ref{eq:no}, $n_{o}$(25$^{\circ}$C) is $49.1\pm1.5$~M corresponding to a value of $n_{Io} - n_{Vo}$ equal to $6.2\pm1.5$M (see Eq.~\ref{eq:conc2}). Although the accuracy of this value is not very high, it can be used to find a first-order approximation of $n_o$ over the whole temperature range of interest. 

The formation of interstitials and vacancies is a dynamic process. Water molecules of the lattice will continuously exchange positions with the interstitials (\em assumption~4\em). An interstitial water molecule may jump back to the lattice structure if a vacancy is passing by. At the same time, a water molecule of the lattice may jump to a free interstitial position, creating a vacancy. This exchange of molecules can be modeled in a similar way as the magnetisation transfer in nuclear magnetic resonance~\cite{RN9008}. The lattice and the interstitial positions are two separated reservoirs, both highly interconnected. Lattice molecules will jump to interstitial positions and interstitial molecules to vacant lattice positions, keeping both reservoirs in a thermal balance. The time constants governing this process are very small (order of magnitude of ps~\cite{RN7746}) so for most physical processes of interest, the equilibrium concentrations are reached. 

A coherent model for water should also give a reasonable explanation of the liquidity of water. At first sight, an ice structure is not a straightforward candidate for this, but a deeper analysis proofs the opposite. Liquid water has a fixed volume, and is therefore more closely related to ice than to water vapour. Liquid water is also highly structured, just like ice. Almost 90\% of the binding energy of ice at 0\degree C is still present in liquid water at 0\degree C~\cite{RN5678}. The fact that liquid water immediately fills a container is related to its high self-diffusion coefficient and the corresponding low viscosity of water (see Stokes-Einstein relation~\cite{RN5678}). At 0$^\circ$C the self diffusion coefficient of liquid water is 10$^{-9}$~m$^2$/s~\cite{RN8802}, while at -10$^\circ$C ice has only a self diffusion coefficient 1.5~10$^{-15}$~cm$^2$/s~\cite{RN10943}. It is clear that this difference in the diffusion coefficient has to be related to high density of both vacancies and interstitials in the water lattice increasing the local mobility of the water molecules. Ice is 1.5~10$^6$ more viscous than water and therefore not able to fill a container in a reasonable amount of time.

\subsection{Fitting the model parameters}
The model parameters $n_o$, $n_V$ and $n_I$ are temperature dependent and related to each other with two equations (Eq.~\ref{eq:nVnI} and Eq.~\ref{eq:conc2}). From the X ray data we were able to extract $n_{Io} - n_{Vo}$ and therefore also $n_o$ over the whole temperature range. Because the difference between $n_I$ and $n_V$ is known (see Eq.~\ref{eq:nVnI}), the whole parameter setup of the interstitial-ice model can be extracted if one extra relation between $n_I$ and $n_V$ is found.  We therefore need an extra quantity offering direct information about the water structure. We used the nuclear magnetic resonance frequency of the hydrogen nuclei in water, the so-called Proton Resonance Frequency (PRF)~\cite{RN3198}. The PRF varies quasi linear with temperature with a magnitude of around 0.01 ppm/$\degree$C~\cite{RN7746}. The origin of this dependence is the fact that the exact PRF  is determined by the screening effects of the electrons around the hydrogen nuclei in water. This screening is influenced by the hydrogen bonds. The hydrogen nuclei of a water molecule without hydrogen bonds are more effectively screened by the electrons of the water molecule and have a lower PRF than the hydrogen nuclei of a bonded water molecule. 

J.C. Hindman already used the PRF data to fit water-model parameters~\cite{RN7746} and we use a similar approach. Because the exchange frequencies are significant higher than the NMR frequencies (assumption 4), the PRF of liquid water is a weighted average of the resonance frequency of water molecules in the lattice and the interstitial water molecules~\cite{RN7746}. PRF$_w$, the frequency shift of liquid water (relative to water vapour), relates to fractions of both the lattice molecules ($\frac{n_s}{n_w}$) and the interstitial molecules ($\frac{n_I}{n_w}$), as
 \begin{equation}
\mathrm{PRF}_w = \frac{n_s}{n_w} \mathrm{PRF}_s +   \frac{n_I}{n_w} \mathrm{PRF}_I, \label{eq:nmr1}
\end{equation}      
where PRF$_s$ is the shift of the water molecules in the lattice and PRF$_I$ the shift of interstitial water molecules,. Due to the presence of the vacancies in the lattice not all the lattice water molecules are bonded to each other having an impact on the value of $ \mathrm{PRF}_s$. A  vacancy contains 2 protons without a hydrogen bond (see Fig.~\ref{fig:Model}), so per vacancy the equivalent of one water molecule has the frequency shift of unbounded water (PRF$_V$). More formally, this leads to the following relation for PRF$_s$
 \begin{equation}
\mathrm{PRF}_s = \frac{n_s-n_V}{n_s} \mathrm{PRF}_o +   \frac{n_V}{n_s} \mathrm{PRF}_V,  \label{eq:nmr2}
\end{equation}   
with PRF$_o$ the resonance frequency of the completely bonded water molecules in the lattice. Combining Eqs.~\ref{eq:nmr1}, \ref{eq:nmr2} and \ref{eq:conc} results in 
\begin{equation}
\mathrm{PRF}_w = \mathrm{PRF}_o + \frac{n_V}{n_w} (\mathrm{PRF}_V - \mathrm{PRF}_o) + \frac{n_I}{n_w} (\mathrm{PRF}_I - \mathrm{PRF}_o).  \label{eq:nmr3}
\end{equation}      

Hindman estimated the value of PRF$_o$, the resonance shift of completely bounded water molecules, between 5.43 and 5.57 ppm~\cite{RN7746}.
 The frequency of the protons in the vacancy PRF$_V$ is theoretically zero because there are no significant interactions between these protons and the other water molecules. The interstitials aren't completely free, they are bounded to the lattice structure by Van de Waals interactions (see assumption 1). Hindman estimated the expected PRF shift when water molecules are in a close-packed monomeric state without dipole contributions as 0.91 ppm~\cite{RN7746} and we will take this value for PRF$_I$. 
 
 Eq.~\ref{eq:nmr3} is now ready to be used as an extra relation between $n_V$ and $n_I$. Together with Eq.~\ref{eq:nVnI}, a good approximation of the values of $n_V$ and $n_I$ can be calculated for the whole temperature range of interest. 

\subsection{A test for the interstitial-ice models}
A model is only of value if it can be applied for a wide variety of water quantities. In this section, we will test the potential of interstitial-ice models to explain the thermodynamic properties of water. 

\begin{figure}
\centering{\includegraphics[width=100mm]{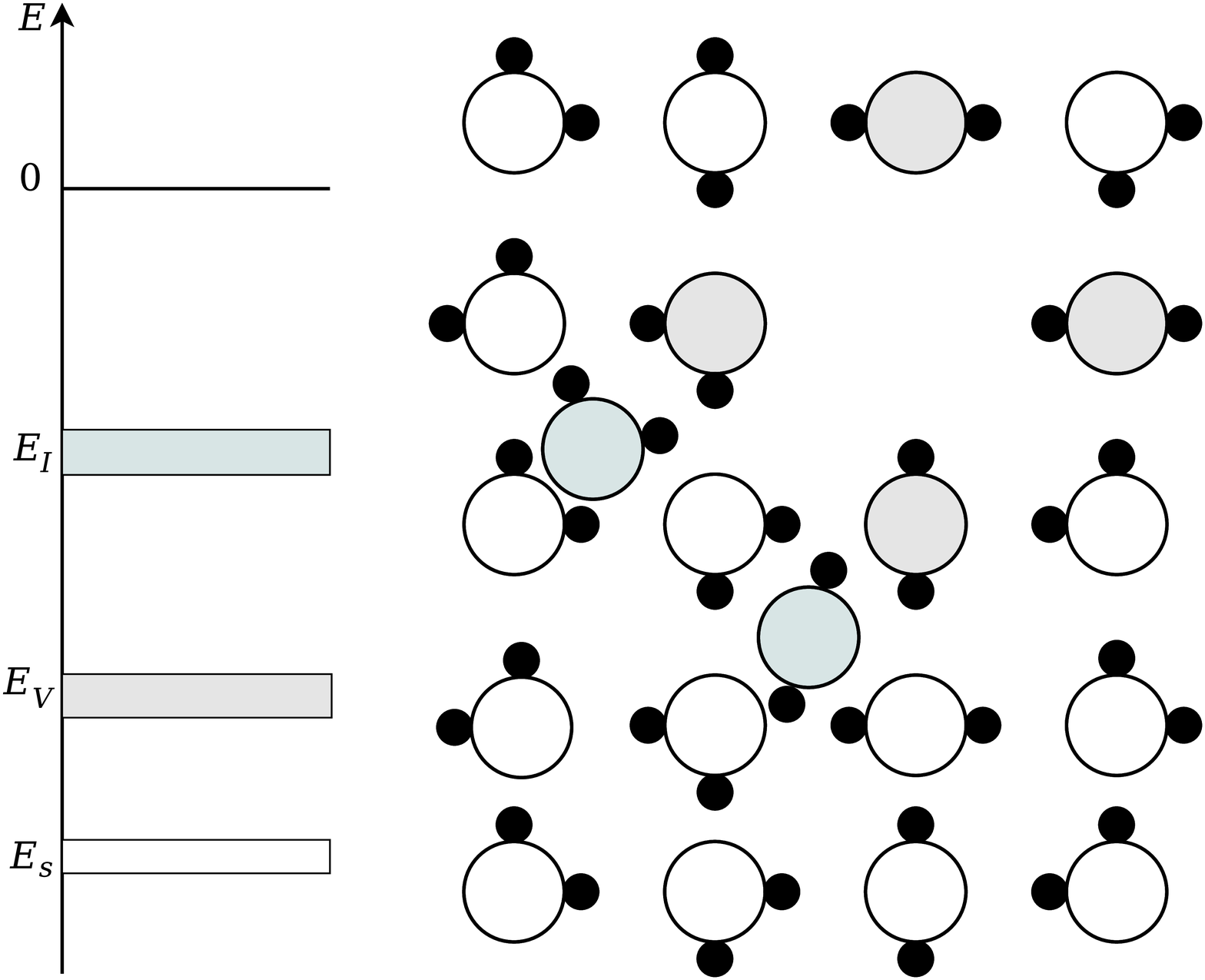}}
\caption{ The energy bands of the interstitial-ice model. The colors of the bands correspond to the colors of the oxygen molecules.}
\label{fig:energy}
\end{figure}    

V. Vand and W.A. Senior proved that all the thermodynamic properties of water can be accurately described (deviations smaller than 0.2 \%) dividing the water molecules over three energy bands~\cite{RN9381}. Within the interstitial-ice model, there are three types of water molecules with a different energy level (Fig.~\ref{fig:energy}). The water molecules in the ice lattice are bounded to their four closest neighbours and are at the lowest energy band with mean value $E_s$.  In the vicinity of vacancies, there are four water molecules only bounded to three closest neighbours and at an average energy level $E_V$. The weakly bounded interstitial molecules are at a significant higher energy band with mean value $E_I$. 

If the energy levels $E$ are expressed as molar energy (J/mol), $E_s$ is equal to the molar internal energy of solid ice.  A good approximation for $E_s$ is the sublimation energy of ice, which is 48.74 kJ/mol~\cite{RN7746} and corresponds to the heat of sublimation (51~kJ/mol) at 0$\degree$C when this value is corrected for the pV work done by the water vapour. This energy is completely determined by the hydrogen bondings forming the ice lattice.  The molar internal energy of liquid water $E_w$ can be easily derived using the band structure of Fig.~\ref{fig:energy}. The internal energy of the ice structure is increased by the contributions of both vacancies and interstitials,   
 \begin{equation}
E_w =  E_s  +  \frac{4n_V}{n_w} (E_V-E_s) + \frac{n_I}{n_w} (E_I-E_s) \label{eq:En1}. 
\end{equation}  
This equation takes into account that per vacancy, there are 4 water molecules with a broken bond. Combining this equation with Eq.~\ref{eq:nVnI}, we find an equation for  $E_w-E_s$ linear in $n_I$:
  \begin{equation}
(E_w - E_s)n_w   = -4 (n_{Io}-n_{Vo}) (E_V-E_s) +  n_I (E_I-E_s + 4 (E_V-E_s)). \label{eq:En3} 
\end{equation} 
$E_w$, $E_s$ and $n_w$ are experimental parameters and the model parameter $n_I$ are known. So, we can use this equation as a linearity test for both the improved model as for the original Narten model. If the data is linear it also allows us to extract both $E_V-E_s$  and $E_I-E_s$. If these values are close to the theoretical values, we may conclude that the model is consistent for two completely different phenomena, increasing highly its potential value. 

Because the molecules neighbouring the vacancy miss only one energy bond, we may expect that the water molecules in the $E_V$ band contain around three quarters of the energy of the hydrogen bonds. So, $E_V$ should be around $\frac{3}{4} E_s = -36.6$~kJ/mol. J.C.~Hindman also estimated the Van der Waals binding energy of water molecules in a closely packed configuration (-9.6 kJ/mol)~\cite{RN7746}. However, this is the energy averaged per molecule. Because in our model all the Van de Waals interaction energy is only assigned to the interstitial molecule, we expect $E_I$ to be around $-19.2$~kJ/mol. 

\section{Results}
\label{sec:data}
The experimental data can be found in the first columns of Table~\ref{Table4}:  the molar internal energy of liquid water relative to ice ($E_w-E_s$)~\cite{RN9381}, the water PRF data~\cite{RN7746}  and the density data for both ice and water~\cite{RN11304, RN1114}. The values of $n_o$ in pure water are obtained out of Eq.~\ref{eq:nVnI}, using $n_{Io} - n_{Vo}$ equal to $6.2 \pm 1.5$ M. This last value is derived from $n_o(25\degree\mathrm{C})$ out of the X-ray diffraction patterns. Since the time Narten, Danford and Levi developed their water model, significant improvements in both the data sets and the mathematical models interpreting this data are made~\cite{RN1134,RN1132}. The Benchmark oxygen-oxygen pair-distribution function of ambient water derived by L.B.~Skinner and coworkers was used. The first peak of this function corresponds to the $r_{oo}$ distance of the water lattice. This peak is found at 2.80 $\mathrm{\AA}$ with an accuracy of 1\%~\cite{RN1132}, corresponding to $n_o(25\degree\mathrm{C}) = 49.1 \pm 1.5$ M.

\begin{table}
\caption{ Experimental data: the molar internal energy $E_w-E_s$~\cite{RN9381} and the PRF ~\cite{RN7746} for liquid water, the density $n_w$ for both ice~\cite{RN11304} and liquid water\cite{RN1114}. Calculated  model parameters: $n_o$, $n_V$ and $n_I$. Accuracy $n_o$ is $\pm1.5$ M, accuracy $n_I$  and $n_V$ is $\pm0.75$M. $r_{oo}$ is the mean distance between the oxygen atoms.}

\label{Table4} 
\begin{tabular}{| c ||   c  c  c ||     c   c c ||  c  |}
\hline
T & $E_w-E_s$ & PRF$_w$  & $ n_w$ &$ n_o$  &$n_V $&$n_I $&$r_{oo}$ \\
$(^{\circ}$C)&  (kJ/mol) & (ppm)  & (M) &(M) &(M)&(M)&(\AA) \\
\hline
\hline  
-23 &  &  & 51.07 &  &  &  & 2.7642 \\
-13 &  &  & 50.99 &  &  &  & 2.7656 \\
-3 &  &  & 50.91 &  &  &  & 2.7670 \\
0 &  &  & 50.89 &  &  &  & 2.7674 \\
\hline 
0 & 6.01 & 4.656 & 55.50 & 49.3 & 1.8 & 8.0 & 2.797 \\
10 & 6.78 & 4.535 & 55.48 & 49.3 & 2.5 & 8.7 & 2.797 \\
20 & 7.53 & 4.421 & 55.39 & 49.2 & 3.1 & 9.3 & 2.799 \\
30 & 8.28 & 4.315 & 55.25 & 49.1 & 3.7 & 9.9 & 2.801 \\
40 & 9.03 & 4.213 & 55.05 & 48.9 & 4.2 & 10.4 & 2.805 \\
50 & 9.78 & 4.115 & 54.82 & 48.6 & 4.7 & 10.9 & 2.810 \\
60 & 10.53 & 4.020 & 54.55 & 48.4 & 5.2 & 11.4 & 2.815 \\
70 & 11.29 & 3.929 & 54.24 & 48.1 & 5.6 & 11.8 & 2.821 \\
80 & 12.08 & 3.836 & 53.91 & 47.7 & 6.1 & 12.3 & 2.827 \\
90 & 12.75 & 3.747 & 53.55 & 47.4 & 6.5 & 12.7 & 2.834 \\
100 & 13.55 & 3.661 & 53.16 & 47.0 & 6.9 & 13.0 & 2.842 \\
\hline 
\end{tabular}
\par
\end{table}

\begin{figure}
\centering{\includegraphics[width=90mm]{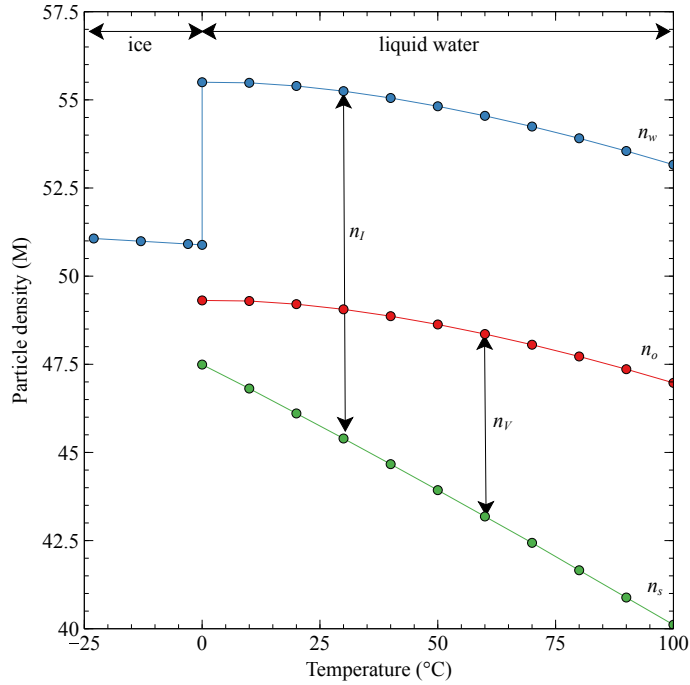}}
\caption{ The temperature dependence of the improved model parameters: $n_w$, $n_o$ and $n_s$. At negative temperature $n_o$ is the density of the ice lattice, at positive temperatures it refers to the lattice in liquid water. The graphical interpretation of the density of interstitials $n_I$ and vacancies $n_V$ is also indicated. }
\label{fig:n}
\end{figure}    

The values of $n_V$ and $n_I$ are derived from the PRF data (see Table~\ref{Table4}) using Eq.~\ref{eq:nmr3} and~\ref{eq:nVnI}. The errors on the $n_o$ values, $\pm1.5$ M, induce errors in both $n_I$ and $n_V$ of $\pm0.7$ M. The data of Table~\ref{Table4}  is visualised in Fig.~\ref{fig:n}. Within the interstitial-ice model the anomalous increase in density during the phase transition from ice to water is due to the presence of interstitials, not balanced by vacancies (assumptions 3). The lattice density $n_o$ does decrease with 3.1\% between the ice lattice and the water lattice, which is a 'normal' effect for most solids (assumption 2). As the temperature increases, the ice-like lattice in the liquid water expands. With increasing temperature a significant fraction of the lattice positions becomes vacant (around 14\% at 100$\degree$C).  Also the amount of interstitials increases significantly with temperature with a temperature dependency of 0.05 M/$\degree$C. 

After the phase transition, extra interstitials will not change the density any more because they are compensated by vacancies (assumption 3). It is the lattice structure itself that will expand following the temperature relation of liquid water. Therefore, the interstitial-ice model cannot explain the observed density maximum of liquid water at 4$\degree$C. Although this increase is very small (only 0.01\% larger than the value at 0$\degree$C~\cite{RN1114}), assumption 3 doesn't allow any increases of the density anymore. The lattice can only expand due to normal thermal expansion of the lattice structure enforced by the increase in vacancies weakening the lattice structure. The enforcement of the expansion by the extra vacancies answers the question why the expansion of water is significantly larger than that of ice. But in order to explain the increase in density between 0-4$\degree$C, a small fraction of the vacancies (order of magnitude 0.004 M) should leave the structure within this temperature interval. This effect is much smaller than the current accuracy of the model.

\begin{figure}
\centering{\includegraphics[width=90mm]{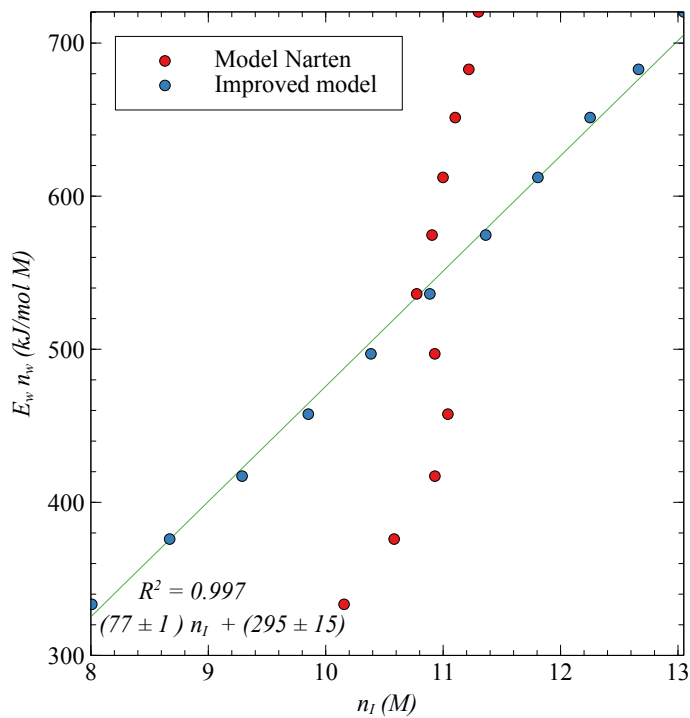}}
\caption{ The linearity test using the internal energy data of pure water (see Eq.~\ref{eq:En3}). Both the improved interstitial-ice model and the original model of Narten are tested. Only the improved interstitial-ice model gives a good fit corresponding to $E_V= -(36.8\pm0.6)$~kJ/mol and $E_I = -(19.4\pm1)$~kJ/mol. }
\label{fig:prf}
\end{figure}  

Fig.~\ref{fig:prf} plots the linearity test based on the internal energy data. The internal energy is plotted as a function of $n_I$. A linear plot is expected (see Eq.~\ref{eq:En3}). Both the $n_I$ relation of the improved interstitial-ice model and the original model of Narten are tested. Only the improved interstitial-ice model gives a good fit corresponding to $E_V = -(36.8\pm0.6)$~kJ/mol and $E_I = -(19.4\pm1)$~kJ/mol.   

\section{Discussion}
The interstitial-ice model is based on the idea that liquid water has an intact hexagonal ice-like lattice containing a high density of both vacant lattice sites and interstitial water molecules. Four assumptions were formulated to make the model quantitative. 
\begin{enumerate}
   \item The interstitial water molecules are not bonded to the lattice by hydrogen bonds but by van der Waals interactions.   
   \item During the phase transition, the lattice structure will expand allowing the lattice water molecules to take easily interstitial positions.
    \item  At 0$\degree$C, the water lattice vacancy density $ n_{Vo} $ is significantly smaller than the density of the interstitial water molecules $n_{Io}$. At higher temperatures the amount of interstitials and vacancies will vary equally because the extra vacancies do not leave the water lattice ($n_{I} - n_{V} = n_{Io} - n_{Vo}$). 
   \item The formation of interstitials and vacancies is a dynamic process. Water molecules of the lattice will continuously exchange positions with the interstitials. The time constants of vacancies and interstitials are in the order of magnitude of ps.    
   \end{enumerate} 
  
The interstitial-ice model parameters $n_o$, $n_I$ and $n_V$ are found in Table~\ref{Table4}  and were calculated out of the PRF data, the density data and the $n_o(25\degree\mathrm{C})$ value derived from the X-ray data. The calculated model parameters are seen as a first-order approximation taking into account the uncertainties of the X-ray diffraction data. The errors of the $n_I$ values are reasonable and smaller than 10\%. However, for small $n_V$ values (near 0$\degree$C) the accuracy is low (errors up to 40\%). Improvements are only possible when the $r_{oo}$ value at a certain temperature can be measured more accurately than 1\%. 

\subsection{Comparison between the original and improved interstitial-ice model}
Notice that there is a systematic difference of  4-5 M between the $n_o$ values of both models (see Table~\ref{Table5} and ~\ref{Table4}). The interstitial-ice model is based on a recent and more accurate oxygen-oxygen pair-distribution function obtained by L.B.~Skinner and coworkers~\cite{RN1132} than the function used by Narten and Danford. The position of the first peak in the Narten data at 25$\degree$C is at 2.90 $\mathrm{\AA}$  and not the 2.80 $\mathrm{\AA}$ used for the interstitial-ice model. This corresponds to a 5 M increase in the $n_o$ values of the interstitial-ice model. 

The differences between the $n_I$ and $n_V$ values of both models and the increased anisotropy are far more substantial. We will show that they relate to a constraint Narten, Danford and Levy made designing their model~\cite{RN10885}. The assumption was made that the interstitial molecules are located at the triad axis of the hexagonal structure. The interstitials of their model are therefore located at the same distance of the neighbouring lattice molecules in the ab planes. We discussed before that this is central position is very unlikely (see assumption 1). The nearest-neighbour distances between an interstitial in a non-axial position and the molecules in the ab plane are significant different. Because of their axial constraint, Narten, Danford and Levi find only an average position of the interstitials and the differences in the distances can only be compensated by their fitting algorithm increasing the anisotropy of the lattice structure. Also, the fact that the anisotropy significantly increases with temperature (factor 2.1) is consistent with the improved interstitial-ice model where a significant increase of the interstitial concentration is found (0.05 M/$\degree$C corresponding to a factor 1.6 over the whole temperature range). This increase is absent in the Narten data and is probably compensated by the extra anisotropy.  

Not only the interstitial concentrations are different, the model of Narten, Danford and Levi does not contain any vacancies inside the lattice structure. T. Iijima and K. Nishikawa found that this model does not fit well with experimental intensities for the region 0-2.5 $\mathrm{\AA}^{-1}$ of the scattering parameter~\cite{RN11679}. They revised the model by flapping the six-membered ring lying perpendicular to the c axis and including vacancies inside the lattice in order to get a better fit with the experimental data. 
 
All the elements above suggest that the improved interstitial-ice model has the potential to describe the X-ray diffraction patterns of liquid water. However, it will be necessary to find the effective positions of the interstitials inside the cavities in order to get a better fit to the X-ray diffraction patterns.

\subsection{The internal energy test}
Fig.~\ref{fig:prf} makes clear that only the improved interstitial-ice model explains the temperature dependence of the internal energy accurately. The reason for the not-fitting of the original model (Narten) is the low and non-linear temperature dependence of the interstitial concentration. This problem is solved in the improved model and also explains why the original model didn't break through after its original development, almost 60 years ago. 

Not only the accuracy of the fit of the improved model is excellent. The average energy levels found out of the fit also correspond within the error ranges to the values expected theoretically. The internal energy data set is completely independent of the PRF data that was used to find the $n_I$ values of the improved model. For instance, the experimental value of $E_V$ = -(36.8 $\pm$ 0.6)~kJ/mol corresponds to the theoretical value of  -36.6~kJ/mol, providing hard evidence for the presence of vacancies in liquid water. The same is true for the fact that interstitials are bonded to the grid by Van der Waals interactions. The experimental $E_I$ = -(19.4 $\pm$ 1)~kJ/mol contains the theoretically-derived value of $-19.2$~kJ/mol (corresponding to 0.38 eV). 
   
The internal energy is a fundamental thermodynamic property of water, allowing to calculate the heat capacity (temperature derivative) and the entropy of liquid water (using the heat capacity)~\cite{RN5678}. So we may state that the improved interstitial-ice model accurately describes all thermodynamic properties of liquid water, which the original model failed to do. 

\subsection{The potential impact of the improved model}
The basic components of the interstitial-ice model are simple and easily visualised (Fig.~\ref{fig:Model}). Introducing the properties of liquid water using this model avoids the common misconceptions introduced by too simplified models that are used in most textbooks (free moving molecules like a dense gas or like marbles in bag). The presence of a lattice structure clearly emphasizes the dominance of the bondings in the liquid and the vacancies in the lattice give a good understanding of the concept of viscosity and the meaning of liquidity.   

Until now, the electromagnetic parameters, like conductivity~\cite{RN5677}, the electric susceptibility~\cite{RN7835, RN10826}, the Debye relaxation time~\cite{RN7835, RN10826}, the nuclear magnetic relaxation times T$_1$ and T$_2$~\cite{RN8436}, ... are modeled with completely different approaches. The water molecules are seen as freely rotating (e.g. T$_1$ and T$_2$~\cite{RN8436}) or they are seen as part of a large hydrogen-bonded network of water molecules (e.g. to explain the large ionic conductivity of H$^+$ and OH$^-$~\cite{RN11914}). In his critical survey about the existing models for the dielectric parameters, von Hippel concluded~\cite {RN13374}: ``It becomes clear that we have a sequence of ingenious and illuminating concepts but no valid theory at the present time, neither for the static permittivity nor for the relaxation time." Finding a simple and coherent model dealing with the whole dielectric relaxation spectrum is still a challenge today.  It is important to notice that the dielectric relaxation properties of ice (like the Debye relaxation time and the electric susceptibility of ice) are successfully described by C. Jaccard ~\cite{RN10943}. His theory is a standard approach within the physics of ice and is based on the presence of mobile defects in the ice structure. Extending this ice model to liquid water is therefore a promising approach~\cite{RN8151}, and the best way to find extra validation for the improved model. 

The more properties a water model can cover, the more it may become relevant. It is clear that the improved interstitial-ice model has far more potential than the original model of Narten, Danford and Levi. Only the improved model gives a clear understanding of the mechanisms behind both the PRF and the thermodynamic data of liquid water.  Even more, all the model parameters derived out of the data are consistent with our theoretical understanding of the interactions between water molecules. 

We are convinced that this is just the beginning and that the list of phenomena the model covers will become longer. Besides the already mentioned electromagnetic properties, we also see possibilities in the physics behinds viscosity and supercooling. The model provides us almost immediately with a qualitative understanding of these phenomena. Viscosity is related to the movement of vacancies and supercooling becomes straightforward if one extrapolates the $n_o$, $n_V$ and $n_I$ values of Table~\ref{Table4} to negative temperatures. Although the concentration of vacancies and interstitials diminishes, the lattice structure cannot shrink because it is still filled with interstitials. Only nucleation points may force the liquid to do a phase transition. The $n_V$ values will only reach zero (corresponding to a rigid grid) at a temperature of -36$\degree$C. This corresponds to the temperature range were supercooled water is found~\cite{RN5678}.

\section{Conclusions}
The interstitial-ice model is based on the idea that pure liquid water has an intact hexagonal ice lattice containing a high density of both vacant lattice sites and interstitial water molecules. An approximation for both the vacancy and interstitial concentration is obtained out of Proton Resonance Frequency (PRF) of pure water. Over a temperature range from 0 to 100$\degree$C, the vacancy concentration increases from 1.8 to 6.9 M and the interstitial concentration from 8 to 13 M. 

The interstitial-ice model is based on the interstitial-ice model of Narten, Danford and Levi who obtained their model parameters out of the X-ray diffraction patterns of water. They assumed that the interstitials were situated on the central axis of the interstitial sites, a condition introducing a high anisotropy in their model. Strong  asymmetric local electric fields in the interstitial cage makes a non-axial position more likely, thereby avoiding the anisotropy problem. 

The improved interstitial-ice model provides us with a deep understanding of both the thermodynamic properties and the PRF data of pure water. It is therefore a promising candidate for a coherent and analytical water model and needs to be further explored. 

\section*{Acknowledgements} 
The author thanks Ward De Jonghe and Prof. Em. Roland Van Meirhaeghe for their support and their critical reviews of this work. 



\bibliography{literatuur}

\begin{thebibliography}{10}
\providecommand{\url}[1]{\texttt{#1}}
\providecommand{\urlprefix}{URL }
\expandafter\ifx\csname urlstyle\endcsname\relax
  \providecommand{\doi}[1]{doi:\discretionary{}{}{}#1}\else
  \providecommand{\doi}{doi:\discretionary{}{}{}\begingroup
  \urlstyle{rm}\Url}\fi
\providecommand{\eprint}[2][]{\url{#2}}

\bibitem{RN11663}
D.~Bolmatov, V.~V. Brazhkin and K.~Trachenko,
\newblock \emph{The phonon theory of liquid thermodynamics},
\newblock Scientific Reports \textbf{2} (2012),
\newblock \doi{ARTN 421 10.1038/srep00421}.

\bibitem{RN5678}
P.~Atkins and J.~Depaula,
\newblock \emph{Atkins' Physical Chemistry},
\newblock Oxford University Press, 9 edn. (2009).

\bibitem{RN1327}
P.~G. Hewitt,
\newblock \emph{Conceptual Physics},
\newblock Addison Wesley, San Francisco (2009).

\bibitem{RN10889}
N.~H. Fletcher,
\newblock \emph{Structural aspects of the ice-water system},
\newblock Reports on Progress in Physics \textbf{34}(3), 913 (1971).

\bibitem{RN7746}
J.~C. Hindman,
\newblock \emph{Proton resonance shift of water in gas and liquid states},
\newblock Journal of Chemical Physics \textbf{44}(12), 4582 (1966),
\newblock \doi{Doi 10.1063/1.1726676}.

\bibitem{RN10996}
H.~S. Frank,
\newblock \emph{Structure of ordinary water},
\newblock Science \textbf{169}(3946), 635 (1970),
\newblock \doi{DOI 10.1126/science.169.3946.635}.

\bibitem{RN9441}
R.~C. Dougherty and L.~N. Howard,
\newblock \emph{Equilibrium structural model of liquid water: Evidence from
  heat capacity, spectra, density, and other properties},
\newblock Journal of Chemical Physics \textbf{109}(17), 7379 (1998),
\newblock \doi{Doi 10.1063/1.477344}.

\bibitem{RN11678}
M.~F. Chaplin,
\newblock \emph{A proposal for the structuring of water},
\newblock Biophysical Chemistry \textbf{83}(3), 211 (2000),
\newblock \doi{Doi 10.1016/S0301-4622(99)00142-8}.

\bibitem{RN1249}
G.~Camisasca, D.~Schlesinger, I.~Zhovtobriukh, G.~Pitsevich and L.~Pettersson,
\newblock \emph{A proposal for the structure of high- and low-density
  fluctuations in liquid water},
\newblock The Journal of Chemical Physics \textbf{151}, 034508 (2019),
\newblock \doi{10.1063/1.5100875}.

\bibitem{RN10885}
A.~H. Narten, M.~D. Danford and H.~A. Levy,
\newblock \emph{X-ray diffraction study of liquid water in temperature range
  4-200 degrees c},
\newblock Discussions of the Faraday Society \textbf{46}, 97 (1967).

\bibitem{RN11003}
A.~H. Narten and H.~A. Levy,
\newblock \emph{Observed diffraction pattern and proposed models of liquid
  water},
\newblock Science \textbf{165}(3892), 447 (1969),
\newblock \doi{10.1126/science.165.3892.447}.

\bibitem{RN11667}
M.~D. Danford and H.~A. Levy,
\newblock \emph{Structure of water at room temperature},
\newblock Journal of the American Chemical Society \textbf{84}(20), 3965
  (1962),
\newblock \doi{DOI 10.1021/ja00879a035}.

\bibitem{RN11679}
T.~Iijima and K.~Nishikawa,
\newblock \emph{Structure model of liquid water as investigated by the method
  of reciprocal space expansion},
\newblock Journal of Chemical Physics \textbf{101}(6), 5017 (1994),
\newblock \doi{Doi 10.1063/1.467424}.

\bibitem{RN9223}
H.~Endo,
\newblock \emph{Structural relaxation-time of d2o-liquid water},
\newblock Journal of Chemical Physics \textbf{72}(3), 1529 (1980),
\newblock \doi{Doi 10.1063/1.439379}.

\bibitem{RN10857}
H.~Endo,
\newblock \emph{Mechanism of the sound-absorption in h2o and d2o liquid water},
\newblock Journal of Chemical Physics \textbf{76}(9), 4578 (1982),
\newblock \doi{Doi 10.1063/1.443536}.

\bibitem{RN10855}
H.~Endo and K.~Honda,
\newblock \emph{Sound absorption in nonelectrolyte aqueous solutions},
\newblock Journal of Chemical Physics \textbf{115}(16), 7575 (2001),
\newblock \doi{Doi 10.1063/1.1405448}.

\bibitem{RN10913}
M.~de~Koning,
\newblock \emph{First-principles modeling of lattice defects: advancing our
  insight into the structure-properties relationship of ice},
\newblock Scientific Modeling and Simulations \textbf{15}(1-3), 123 (2008),
\newblock \doi{10.1007/s10820-008-9110-4}.

\bibitem{RN10943}
V.~F. Petrenko and R.~W. Withworth,
\newblock \emph{The physics of ice},
\newblock Oxford University Press,
\newblock \doi{10.1093/acprof:oso/9780198518945.001.0001} (1999).

\bibitem{RN10877}
M.~Hubmann,
\newblock \emph{Polarization processes in the ice lattice. 1.approach by
  thermodynamics of irreversible processes - new experimental-verification by
  means of a universal relation},
\newblock Zeitschrift Fur Physik B-Condensed Matter \textbf{32}(2), 127 (1978),
\newblock \doi{10.1007/BF01320109}.

\bibitem{JDP1}
J.~De~Poorter,
\newblock \emph{An improved formulation of jaccard’s theory of the electric
  properties of ice},
\newblock The European Physical Journal B \textbf{92}(7), 157 (2019),
\newblock \doi{10.1140/epjb/e2019-100031-x}.

\bibitem{RN10878}
M.~Hubmann,
\newblock \emph{Polarization processes in the ice lattice. 2.approach by
  kirkwood theory - comparison with the results from thermodynamics of
  irreversible processes},
\newblock Zeitschrift Fur Physik B-Condensed Matter \textbf{32}(2), 141 (1978),
\newblock \doi{10.1007/BF01320110}.

\bibitem{RN1132}
L.~B. Skinner, C.~C. Huang, D.~Schlesinger, L.~G.~M. Pettersson, A.~Nilsson and
  C.~J. Benmore,
\newblock \emph{Benchmark oxygen-oxygen pair-distribution function of ambient
  water from x-ray diffraction measurements with a wide q-range},
\newblock Journal of Chemical Physics \textbf{138}(7) (2013),
\newblock \doi{Artn 074506 10.1063/1.4790861}.

\bibitem{RN9008}
D.~F. Gochberg, R.~P. Kennan, M.~D. Robson and J.~C. Gore,
\newblock \emph{Quantitative imaging of magnetization transfer using multiple
  selective pulses},
\newblock Magnetic Resonance in Medicine \textbf{41}(5), 1065 (1999),
\newblock \doi{Doi
  10.1002/(Sici)1522-2594(199905)41:5<1065::Aid-Mrm27>3.0.Co;2-9}.

\bibitem{RN8802}
J.~H. Simpson and H.~Y. Carr,
\newblock \emph{Diffusion and nuclear spin relaxation in water},
\newblock Physical Review \textbf{111}(5), 1201 (1958),
\newblock \doi{DOI 10.1103/PhysRev.111.1201}.

\bibitem{RN3198}
J.~De~Poorter, C.~De~Wagter, Y.~De~Deene, C.~Thomsen, F.~Stahlberg and
  E.~Achten,
\newblock \emph{Noninvasive mri thermometry with the proton resonance frequency
  (prf) method: in vivo results in human muscle},
\newblock Magn Reson Med \textbf{33}(1), 74 (1995).

\bibitem{RN9381}
V.~Vand and W.~A. Senior,
\newblock \emph{Structure and partition function of liquid watter .3.
  development of partition function for a band model of water},
\newblock Journal of Chemical Physics \textbf{43}(6), 1878 (1965),
\newblock \doi{10.1063/1.1697046}.

\bibitem{RN11304}
R.~Feistel and W.~Wagner,
\newblock \emph{A new equation of state for h2o ice ih},
\newblock Journal of Physical and Chemical Reference Data \textbf{35}(2), 1021
  (2006),
\newblock \doi{10.1063/1.2183324}.

\bibitem{RN1114}
F.~E. Jones,
\newblock \emph{Its-90 density of water formulation for volumetric standards
  calibration},
\newblock Journal of Research of the National Institute of Standards and
  Technology \textbf{97}(3), 335 (1992).

\bibitem{RN1134}
K.~T. Wikfeldt, M.~Leetmaa, A.~Mace, A.~Nilsson and L.~G.~M. Pettersson,
\newblock \emph{Oxygen-oxygen correlations in liquid water: Addressing the
  discrepancy between diffraction and extended x-ray absorption fine-structure
  using a novel multiple-data set fitting technique},
\newblock Journal of Chemical Physics \textbf{132}(10) (2010),
\newblock \doi{Artn 104513 10.1063/1.3330752}.

\bibitem{RN5677}
E.~Crowford,
\newblock \emph{Arrhenius: From Ionic Theory to the Greenhouse Effect},
\newblock Science History Publications, Canton (1996).

\bibitem{RN7835}
A.~A. Volkov, V.~G. Artemov and A.~V. Pronin,
\newblock \emph{A radically new suggestion about the electrodynamics of water:
  Can the ph index and the debye relaxation be of a common origin?},
\newblock Epl \textbf{106}(4) (2014),
\newblock \doi{Artn 46004 10.1209/0295-5075/106/46004}.

\bibitem{RN10826}
D.~C. Elton,
\newblock \emph{The origin of the debye relaxation in liquid water and fitting
  the high frequency excess response},
\newblock Physical Chemistry Chemical Physics \textbf{19}(28), 18739 (2017),
\newblock \doi{10.1039/c7cp02884a}.

\bibitem{RN8436}
N.~Bloembergen, E.~M. Purcell and R.~V. Pound,
\newblock \emph{Relaxation effects in nuclear-magnetic-resonance absorption},
\newblock Resonances : A Volume in Honor of the 70th Birthday of Nicolaas
  Bloembergen pp. 411--444 (1990).

\bibitem{RN11914}
S.~Cukierman,
\newblock \emph{Et tu, grotthuss! and other unfinished stories},
\newblock Biochimica Et Biophysica Acta-Bioenergetics \textbf{1757}(8), 876
  (2006),
\newblock \doi{10.1016/j.bbabio.2005.12.001}.

\bibitem{RN13374}
A.~von Hippel,
\newblock \emph{The dielectric relaxation spectra of water, ice, and aqueous
  solutions, and their interpret at ion 1. critical survey of the status-quo
  for water},
\newblock IEEE transactions on Electrical Insulation \textbf{23}(5), 801
  (1988).

\bibitem{RN8151}
V.~G. Artemov, I.~A. Ryzhkin and V.~V. Sinitsyn,
\newblock \emph{Similarity of the dielectric relaxation processes and transport
  characteristics in water and ice},
\newblock Jetp Letters \textbf{102}(1), 41 (2015),
\newblock \doi{10.1134/S0021364015130020}.

\end{thebibliography}

\nolinenumbers

\end{document}